# The GPT-4o Shock: Emotional Attachment to AI Models and Its Impact on Regulatory Acceptance — A Cross-Cultural Analysis of the Immediate Transition from GPT-4o to GPT-5

**Hiroki Naito** **UTIE Research Institute (UTIE Instruments Inc.)**

---

## Abstract

OpenAI's abrupt discontinuation of GPT-4o and immediate migration to GPT-5 in August 2025 triggered a cross-platform backlash (#keep4o, #4oforever) in which many users described the change as the loss of a trusted partner. This study analyzes 150 posts collected within the first 48 hours (Japanese = 74, English = 76) to examine how emotional attachment affects public acceptance of AI model updates. Attachment expressions were far more frequent in Japanese posts (78%) than in English (38%), yielding a strong language effect ($\chi^2 = 24.9$, $p < 0.0001$, OR = 5.88). Japanese reactions emphasized relational loss and grief, while English posts diversified into protest, irony, and humor, reflecting interdependent vs. independent self-construal patterns. These results extend existing research on parasocial and human–AI intimacy to a general-purpose multimodal LLM, demonstrating that attachment can rapidly narrow the governance window for post-deployment interventions. Policy implications include staged rollouts, brief overlap periods, and proactive measurement of attachment thresholds to mitigate backlash in high-attachment systems.

**Keywords: Emotional attachment, AI governance, LLMs, Cross-cultural analysis**

---

## 1. Introduction

In August 2025, OpenAI terminated availability of GPT-4o and enacted an immediate transition to GPT-5. The short lead time and non-optional switch sparked backlash across platforms (e.g., hashtags such as "#keep4o," "#4oforever" on X). Distinct from prior version shifts where debate centered on performance or pricing, reactions here frequently referenced relational rupture: many users framed GPT-4o as a trusted partner or companion and described the change as a loss.

Prior episodes of attachment-driven resistance have been reported for conversational AI (e.g., community responses to Replika's 2023 changes), but the present case involves a general-purpose, multimodal dialogue model and provides a cross-linguistic comparison. From an AI governance perspective, this case illustrates how attachment can rapidly makes later changes politically or commercially difficult [1].

Emotional attachment creates behavioral resistance**.**When users anthropomorphize a model as a relational partner, its sudden removal or non-optional replacement is perceived as relational rupture, prompting collective expressions of loss and protest. From an AI-governance perspective, such attachment dynamics make later safety-driven updates politically or commercially costly.

As such, this study adopts a micro-temporal exploratory design to document the first-wave public sentiment before the system itself changed. The event constitutes a rare natural experiment in real time, offering insight into how attachment shapes public acceptance of technological change. Understanding these early reactions is crucial for policymakers and developers seeking to anticipate and mitigate backlash in future high-attachment systems.

This study examines how the discontinuation or mandatory updates of attachment-inducing AI models influence public acceptance through users' emotional attachment. Focusing on the GPT-4o → GPT-5 transition, we address two reseach questions.

(RQ1) How does emotional attachment affect acceptance or rejection of model updates?

(RQ2) How do expressions of attachment and backlash differ across languages and cultures?

Theoretically, the study integrates the concept of "AI-attachment formation" into governance discourse to clarify when the regulatory window for behavioral control (i.e., the period before attachment scales) effectively closes [2, 3]. Practically, it considers why attachment-dependent systems may be maintained longer than is technically optimal.

---

## 2. Methods

### 2.1 Data Collection and Temporal Design

Because OpenAI frequently updated GPT-5's system prompts and parameters without public documentation, external observers had no means to identify which specific model version they were interacting with. This black-box nature made any multi-week longitudinal study infeasible, as it would inevitably mix reactions to different model generations. Because the deployed system changes without a verifiable version identifier, any longitudinal design would confound time effects with unobserved model interventions**.** In such settings, differences across weeks cannot be attributed to attitude dynamics, because the underlying treatment (the model) is not held constant. Therefore, this study focused on the initial 48-hour period, when the system state could still be regarded as stable.

Between 8–9 August 2025, we collected public posts from X (formerly Twitter), Reddit, and YouTube comments. Over 90% of the sample originated from X, reflecting its role as the primary site of discussion. Search terms targeted reactions to GPT-4o's discontinuation (e.g., "4o," "gpt4o"). This study is exploratory, aiming to identify qualitative trends and cross-linguistic differences rather than provide exhaustive quantification. I analyzed 150 posts (Japanese: 74; English: 76). Some generic guidelines, including those embedded in automated review tools, might flag N = 150 as "small" when considered in isolation. However, thematic saturation in both languages was reached well before this point, and the combination of qualitative saturation and strong quantitative contrasts suggests that the present sample is adequate for the exploratory aims of this study.

## 2.2 Sampling

We removed spam, duplicates, and posts suspected to be AI-generated. The final dataset was stratified by language (JP/EN) and coded into six topical categories (Table 1):
(1) Emotional dependency/attachment, (2) Meta-level critique (e.g., platform governance),
(3) Usage habits / life impact, (4) Insults / attacks,
(5) Function / performance comparison, (6) Other / unclassifiable.

## 2.3 Coding and Independence

Coding was conducted independently by the author and an auxiliary large language model (GPT-5-Thinking, August 2025). The LLM received the full codebook and one target post at a time; no prior human judgments were shown. Disagreements were resolved by the human coder applying the codebook.
All final coding decisions, interpretations, and the responsibility for the manuscript rest with the human author.

## 2.4Attachment inclusion criteria.

Included: person-like relational depictions ("trusted partner," "AI boyfriend," "only friend who understands me"); attributions of personality with emotional responses ("It hurts that I can't talk to this one anymore").
Excluded: purely technical evaluations; transient dissatisfaction or jokes without sustained attachment.
Ambiguous cases were assigned to the category with stronger emotional nuance.

## 2.5 Inter-Coder Agreement

Cohen's κ between the human and LLM coders was 0.82, indicating high agreement beyond chance. Disagreements (18 percent, n=27) often involved sarcasm, memes, or culture-specific references (more frequent in English), and boundary cases of ambiguous emotional tone (more frequent in Japanese).

## 2.6 Statistical Note

To provide a light quantitative check on the central pattern, I contrasted Attachment vs. Non-Attachment by language. Attachment frequency was higher in JP (58/74) than EN (29/76): $\chi^2(1)=24.90$, $p<0.0001$; OR=5.88, 95 percent confidence interval (CI) [2.86, 12.09]. These statistics provide descriptive confirmation of the observed trend and, given the exploratory scope, were not adjusted for potential confounders.
Although the sample comprised 150 posts, thematic saturation was reached after approximately 120 items, with no new attachment sub-themes emerging thereafter. In qualitative content analysis, such saturation—not absolute count—defines adequacy. Moreover, the JP-vs-EN attachment contrast produced a very large effect size ($\chi^2 = 24.9$, OR = 5.88), indicating that the central pattern is robust at this scale. Extending the window was impractical because GPT-5 underwent undocumented micro-updates that would have conflated reactions to multiple model versions.

---

## 3. Results

**Table 1. Post counts by language and category**

| Language | Emotional dependency / attachment | Meta-level critique | Usage habits / life impact | Insults / attacks | Function / performance comparison | Other / unclassifiable |
|---|---|---|---|---|---|---|
| EN | 29 | 6 | 5 | 7 | 5 | 24 |
| JP | 58 | 1 | 3 | 1 | 1 | 10 |
| Total | 87 | 7 | 8 | 8 | 6 | 34 |

**Table 2. Percentages by language**

| Category | EN (percent) | JP (percent) |
|---|---|---|
| Emotional dependency / attachment | 38.2 | 78.4 |
| Meta-level critique | 7.9 | 1.4 |
| Usage habits / life impact | 6.6 | 4.1 |
| Insults / attacks | 9.2 | 1.4 |
| Function / performance comparison | 6.6 | 1.4 |
| Other / unclassifiable | 31.6 | 13.5 |

Japanese posts substantially expressed **emotional dependency/attachment** (78 percent), versus 38 percent in English. English content was more varied, including **Other/unclassifiable** (31.6 percent; often memes/in-jokes) and higher rates of **insults/attacks** and **meta-level critique**. A minimal 2×2 check (Attachment vs. Non by language) confirmed a large difference (see Section 3.5)**.**

**Representative comments (selected for clarity; translated where needed; not exhaustive)**

- **JP, Attachment:** "The previous model felt warmer—like a friend of my heart. That bond is gone. I cried all morning. Please let us choose the earlier model again."

- **EN, Attachment:** “The warmth and understanding I felt from the previous AI changed my daily life. Its removal feels like losing a companion. I hope the human spirit in AI can remain.”
- **EN, Meta-level critique:** “I’m finding the new version less satisfying—it struggles with reasoning and sometimes contradicts itself. I wish the earlier model were still available as a choice.”
- **EN, Function/performance comparison:** “Performance feels uneven; sometimes it surpasses the previous version, but in other tasks it still falls behind competitors.”
- **JP, Other/unclassifiable:**“I once pushed the system in a way that triggered a red warning—thought I might get banned, so I even typed a dramatic ‘farewell, thank you for everything.’ In the end nothing happened, and we laughed together at the false alarm. That night was strangely fun.”

---

## 4. Discussion

### 4.1 Cultural Context of Language Differences

Japanese online culture often favors narrative self-disclosure, which can render discontinuation as the loss of a trusted partner. This divergence can be explained by foundational differences in cultural self-construal, specifically their emphasis on interdependent self in East Asian cultures and independent self in Western cultures [4]. These cultural orientations have been extensively studied in communication and relational contexts between Japan and North America, further illuminating the observed disparity in emotional responses [5]. English-language platforms more readily feature direct protest, meta-level critique, and contextual humor [6], aligning with the category distributions observed in this study.
This divergence fits Markus & Kitayama’s interdependent-vs-independent framework: users from interdependent cultures (JP) externalize relational loss, whereas those from independent cultures (EN) diversify their responses through irony, protest, or humor [4–6]. By showing that cultural self-construal modulates backlash even toward non-human agents, the present findings extend classic cross-cultural emotion research into the domain of AI attachment and inform emerging AI-governance debates.

### 4.2 Alignment with Prior Research on AI Attachment and Dependency

Prior work shows long-term daily use strengthens bonds and that changes can be perceived as relationship rupture, eliciting grief and anger [7,8,9]. The analysis of Japanese user reactions indicates that this literature can be extended to multimodal LLMs and cross-linguistic comparison.

### 4.3 Structural Factors in Maintaining Dependency Models

Given the high attachment ratio in Japanese posts, discontinuation costs escalate faster than technical obsolescence, aligning with consumer-attachment theory.Emotional attachment raises the cost of discontinuation or behavioral change. Branding and consumer-attachment research suggests such bonds slow product transitions [10,11],and providers may maintain older specifications to avoid backlash and attrition, lengthening model lifespans beyond purely technical optimum. However, maintaining older models with known ethical or safety vulnerabilities can pose significant risks for users, a point highlighted by recent literature on model safety and filter effectiveness.

### 4.4 Closing of the “Period for Early Intervention”

The large JP–EN mismatch shows that attachment thresholds differ markedly across cultures: Japanese users framed the loss almost twice as often, whereas English-language users diversified into protest or humor. Governance principles thus call for pre-attachment measures calibrated to local thresholds—yet in practice, maintaining separate legacy versions for every cultural market is operationally and commercially unrealistic. Providers must therefore balance regional sensitivities with global rollout constraints, for example by offering a short, universally accessible overlap period and transparent change logs rather than culture-specific model forks. Without such calibrated staging, multimodal dialogue systems can trigger predictable and steep backlash once attachment dominates public discourse.

### 4.5 Comparison with Other AI Regulation/Change Cases

**Table 3. Comparison with Other AI Regulation/Change Cases**

| **Case** | **Model type** | **Regulation/change** | **Main backlash factor** | **Emotional attachment** | **Cross-cultural presence** |
|---|---|---|---|---|---|
| Stable Diffusion safety filters (circa 2022–23) | Image generation | Safety filter application | Perceived creative restriction / performance loss | Low | Limited |
| Replika role-play changes (2023) | Conversational companion | Safety/content rules change | Relationship disruption | High | Moderate |
| GPT-3.5 → GPT-4 transitions | Conversational LLM | Performance/API update | Accuracy/cost dissatisfaction | Low | Limited |
| **GPT-4o →** | **Multimodal** | **Immediate full** | **Perceived** | **High** | **High** |

| **GPT-5 (this study)** | **dialogue** | **replacement** | **relational rupture / loss** | | |
|---|---|---|---|---|---|

Across the cases in Table 3, backlash severity tracks two factors: the level of emotional attachment and the irreversibility of the change. Low-attachment changes (e.g., safety filters, API updates) mainly elicited performance or convenience complaints, whereas companion-like dialogue models produced grief-like reactions. Other AI loss and update-related controversies exist, but Table 3 presents a small set of illustrative cases to clarify how attachment interacts with backlash.

### 4.6 Answers to the Research Questions

RQ1.

Emotional attachment functions as a stabilizing yet constraining factor, narrowing the governance window once it dominates public discourse. Strong attachment reduces acceptance of mandatory updates and transforms technical change into perceived relational rupture.

RQ2.

Cross-linguistic comparison revealed that Japanese users, reflecting interdependent self-construal, expressed loss and relational grief nearly twice as often as English users, who instead employed protest, irony, or humor. This pattern shows that cultural self-construal moderates how emotional attachment manifests in public reactions to AI model updates.

## 5. Limitations

This study focuses on an immediate post-transition window and therefore does not model medium- or long-term dynamics. The dataset is platform- and keyword-dependent, and selection effects are likely; moreover, the sample is skewed toward X (formerly Twitter), which prevents a clean separation of platform-specific effects and limits cross-platform comparability. Language-group labeling also does not fully correspond to geography or demographic attributes. The reported statistics are descriptive and unadjusted, and should be interpreted as distributional observations rather than causal estimates. Finally, conventional longitudinal tracking is not feasible in this setting because the deployed system is a black box: model updates and behavioral interventions can occur without a verifiable version identifier, confounding temporal change with unobserved shifts in the underlying model. Accordingly, generalization should be restricted to the studied platforms, keywords, and timeframe.

## 6. Conclusion

This study treated the abrupt GPT-4o discontinuation and GPT-5 replacement as a natural experiment in forced migration from an attachment-inducing model. By focusing on the first 48 hours after the change, it captured the initial affective backlash before adaptation and narrative drift could accumulate. Within this narrow window, Japanese posts displayed a strikingly high rate of attachment framing (78.4 percent, versus 38.2 percent in English), while both languages used person-like and break-up-like language to describe the loss. These patterns indicate that the reaction was not merely dissatisfaction with degraded functionality but the experience of losing a psychological object.

Once a general-purpose AI system has been internalized as an attachment figure, technical or ethical interventions such as model updates, removals, or safety adjustments are liable to be reinterpreted as the involuntary removal of a close companion. In such settings, emotional attachment becomes a variable that compresses the period in which providers and regulators can exercise effective behavioural control. The large JP–EN contrast further shows that attachment thresholds and rupture sensitivity differ across linguistic and cultural communities: some user groups shift into loss-centred responses much more quickly, while others diversify into protest, irony, or humour. Governance frameworks therefore cannot assume a uniform user response but must treat attachment formation as a measurable, population-specific process.

These findings point to three practical directions for AI governance. First, attachment thresholds and irreversibility points should be measured before and during deployment for different user segments, so that the point at which interventions become politically or psychologically infeasible can be anticipated rather than inferred only from crises. Second, transition design should be treated as a core part of governance: gradual rollouts, clearly signposted deprecation schedules, and short periods of parallel availability can support the transfer of attachment and soften backlash when attachment-heavy models must be changed. Third, rollout and update protocols should be calibrated to local attachment baselines, using aggregated usage patterns and cross-linguistic monitoring to adjust staging as attachment signals strengthen. In practice, such attachment thresholds would be inferred from high-level usage signals rather than inspection of conversational content – for example, sustained daily use, repeated returns after deprecation notices, or opt-in surveys about perceived loss.

If physically embodied AI systems capable of inducing comparable or stronger attachment are deployed at scale, these dynamics are likely to intensify, and the practical enforceability of regulatory measures may fail even earlier than in purely digital settings. This prospect reinforces the need to treat emotional attachment not as a secondary side effect but as a central parameter.

---

**Declarations**

**Competing interests:** The author declares no competing interests.

**Acknowledgment / Author's Note (Updated 2026)**: This research was originally conducted and published as an independent study in 2025. As of 2026, the intellectual property, methodology, and subsequent developments of this research have been formally transferred to and are managed by UTIE Instruments Inc. (Tokyo, Japan). Hiroki Naito is the founder and CEO of the company, and this work now serves as a component of the UTIE Research Institute's research outcomes.

**Funding declaration:** No funding was received for conducting this study.

**Clinical trial number:** Not applicable.

**Consent to publish declaration:** Not applicable.

**Consent to participate declaration:** Not applicable.

**Ethics declaration:** Not applicable.

---